\documentclass[aps,pra,superscriptaddress,amsmath,amssymb,preprintnumbers,floatfix,showpacs,11pt]{revtex4}

\usepackage{amssymb}
\usepackage{epsfig}

\begin{document}
\title{ The revival-collapse phenomenon in the   quadrature field
components of the two-mode multiphoton Jaynes-Cummings model
 }
\author{ Faisal A. A. El-Orany
 }
 \affiliation{ Department of Mathematics  and Computer Science,
Faculty of Science, Suez Canal University,
 Ismailia, Egypt}

\date{\today}

\begin{abstract}
In this paper we consider a system consisting of a two-level atom
in an excited state interacting with two modes of a radiation
field prepared initially in $l$-photon coherent states. This
system is described by two-mode multiphoton (, i.e., $k_1, k_2$)
Jaynes-Cummings model (JCM). For this system we investigate the
occurrence of the
 revival-collapse phenomenon (RCP) in
the evolution of the single-mode, two-mode, sum and difference
quadrature squeezing. We show that there is  a class of states for
which all these types of squeezing exhibit RCP similar to that
involved in the corresponding atomic inversion. Also  we show
numerically that the single-mode squeezing of the first mode for
 $(k_1,k_2)=(3,1)$ provides RCP similar to that
of the atomic inversion of the case $(k_1,k_2)=(1,1)$, however,
sum and difference squeezing give partial information on that
case. Moreover, we show that single-mode, two-mode and sum
squeezing for the case $(k_1,k_2)=(2,2)$ provide information on
the atomic inversion of the single-mode two-photon JCM. We derive
the rescaled squeezing factors giving accurate information on the
atomic inversion for all cases. The consequences of these results
are that the homodyne and heterodyne detectors can be used to
detect the RCP for the two-mode JCM.

\end{abstract}

 \maketitle

\section{Introduction}
 The simplest model in quantum optics is the Jaynes-Cummings model
 (JCM), in which a radiation field interacts with a single two-level
 atom \cite{jay1}. This system
  has  become experimentally realizable with
the Rydberg atoms in high-$Q$ microwave cavities (, e.g., see \cite{remp}).
In the framework of the rotating wave approximation  many of
interesting effects have been reported for JCM.
The most important phenomenon  is the revival-collapse phenomenon (RCP),
which occurs in the
evolution of the atomic inversion $\langle\hat{\sigma}_{z}(T)\rangle$, i.e.
 instead of displaying  steady Rabi oscillations in the
case of a classical field coupled to the atom \cite{allen}, there is an initial
collapse of these oscillations followed by regular revivals that slowly
become broader and eventually overlap \cite{eber}.
This indicates that  RCP is a pure quantum mechanical effect \cite{zaber1}.
For more details about the RCP of the JCM the reader can consult,
e.g. \cite{eber}.

The JCM has been extended to include multimode fields, e.g.
\cite{fas3,fas1,fas2}, multilevel atoms \cite{multil} and
multiatom interactions \cite{multia}. Two-mode JCM (TJCM) has
taken a considerable interest and studied from different points of
view, e.g. \cite{fas3,fas2,chr,two}. RCP of the TJCM is rather
complicated compared to that of the single-mode JCM in the sense
that the revival series is compact and each revival is followed by
secondary revival. Furthermore, for strong intensities the
locations of the revival patterns in the time interaction "domain"
are independent of the intensities  (see Fig. 1). Such behaviour
has been partially explained in \cite{chr}, however, an
investigation for the occurrence of the secondary revivals is
given
 in \cite{fas3}.

Quite recently a new technique is developed for discussing how within the
single-mode multiphoton JCM the RCP of the atomic inversion of the
standard (, i.e. single-photon) JCM
is manifested in the evolution of the quadrature squeezing of
the field \cite{faisal2}.
Two approaches have been adopted for such analysis, namely,
natural  and numerical-simulation approaches.
For natural approach it has been shown that there
is a class of states whose squeezing factors  can directly
include information
on the corresponding atomic inversion.
However, the numerical-simulation  approach has been given to show that
the evolution of the quadrature squeezing of
the three-photon JCM  reflects the RCP involved in
the $\langle\hat{\sigma}_{z}(T)\rangle$ of the single-photon JCM for the same
initial field state.
  Moreover, we have deduced a general form for the
higher-oder squeezing factor, which can give information on the atomic
inversion of the single-photon JCM \cite{fais1} and two-photon JCM \cite{fais2}.

In this paper we apply the technique given in \cite{faisal2} to TJCM for
investigating the occurrence of the RCP in the quadrature squeezing and how can be
connected with the atomic inversion. For convenience we assume that the radiation fields
 are initially prepared in  $l$-photon coherent states \cite{jex1,jex2,green}
 and the atom is in an excited atomic state.  Needless to say that
the situation for the TJCM is more complicated than that of the single-mode JCM.
For instance, there are different  types  of quadrature squeezing such as
single-mode, two-mode, sum and difference squeezing. Moreover, the strong
entanglement between the two bosonic systems over the atomic system making
the investigation is rather complicated. In spite of these difficulties we
have obtained many interesting results, e.g. for all types of squeezing there
is a class of states for which squeezing factors  can directly give
 the corresponding atomic inversion. Additionally,
using numerical technique we have shown that
 when  $k_1+k_2=4$ (cf. (\ref{6})) the $Y$-quadrature squeezing factor
of the particular types can provide
 RCP similar to that exhibited in the evolution of the atomic inversion of the standard
 (, i.e., $k_1=k_2=1$) TJCM
or single-mode two-photon JCM based on the values of $k_j$.
We have to stress that the
nonclassical squeezing for TJCM has been
 studied by several authors, e.g., see \cite{mah}, and it will not
be considered in the present paper. Finally, the results given
here and in \cite{faisal2,fais1,fais2} show that the RCP occurred
in $\langle\hat{\sigma}_{z}(T)\rangle$ can be detected using
techniques similar to those used for quadrature squeezing, e.g.
 homodyne detector \cite{homo}, nonlinear
homodyne detector \cite{wilk} and multiport homodyne detector \cite{walk}.
It is worth mentioning that in cavity QED, the homodyne detector
technique has been applied to the single Rydberg atom and one-photon field
for studying  the field phase evolution of the regular  JCM \cite{haroc}.
Quite recently similar  setup is given for
induced measurement  and quantum computation with atoms in optical
cavities \cite{anders}.
Moreover, the progress in both of the trapped ions \cite{Lei} and micromaser
\cite{mas1} is promising to produce the phenomena discussed in the paper.
We conclude this part by drawing the attention to that
the first experimentally observed squeezed states are of the
two-mode type \cite{slu}.

The paper is prepared in the following order.
In section 2 we give the basic relations and equations including the
model and the definition of squeezing. In sections 3--5 we investigate
single-mode squeezing, two-mode squeezing and
sum-difference squeezing, respectively. In section 6 the main
conclusions are summarized.

\section{Basic relations and equations}
In this section we give the basic relations and equations, which will be used
throughout the paper. Precisely, we write down the Hamiltonian of the system under
consideration, its wave function and the definition of quadrature squeezing.

The Hamiltonian controlling the TJCM in the rotating wave
approximation is \cite{two}:
\begin{equation}
\frac{\hat{H}}{\hbar}=
\omega_{1}\hat{a}_{1}^{\dagger}\hat{a}_{1}+
\omega_{2}\hat{a}_{2}^{\dagger}\hat{a}_{2}+
\frac{1}{2}\omega_{a}\hat{\sigma}_{z}+
g(\hat{a}_{1}^{ k_{1}}\hat{a}_{2}^{k_{2}}\hat{\sigma}_{+} +
\hat{a}_{1}^{\dagger k_{1}}\hat{a}_{2}^{\dagger k_{2}}\hat{\sigma}_{-}),
 \label{6}
\end{equation}
where $\hat{\sigma}_{\pm}$ and $\hat{\sigma}_{z}$ are the Pauli spin
operators;
$\omega_{j}, (j=1,2)$ and $\omega_{a}$ are the frequencies of
the cavity modes $\hat{a}_j$ and the atomic frequency, respectively;  $g$ is the atom-field coupling
constant and $k_{j}$ is the transition parameter of the $j$th mode.
The derivation of the Hamiltonian (\ref{6}) from the first principle is given in
\cite{avia}.

We restrict the investigation to the exact resonance
 case  $k_{1}\omega_{1}+k_{2}\omega_{1}=\omega_{a}$.
For evaluating  the dynamical state of (\ref{6}) we define
 two  operators $\hat{F}_{1}$ and $\hat{F}_{2}$ as
\begin{equation}
\hat{F}_{1}=
\omega_{1}\hat{a}_{1}^{\dagger}\hat{a}_{1}+
\omega_{2}\hat{a}_{2}^{\dagger}\hat{a}_{2}+
\frac{1}{2}\omega_{a}\hat{\sigma}_{z},\qquad
\hat{F}_{2}=
g(\hat{a}_{1}^{ k_{1}}\hat{a}_{2}^{k_{2}}\hat{\sigma}_{+} +
\hat{a}_{1}^{\dagger k_{1}}\hat{a}_{2}^{\dagger k_{2}}\hat{\sigma}_{-}). \label{7}
\end{equation}
It is easy to prove that $\hat{F}_{1}$ and $\hat{F}_{2}$ are constants of motion.
This fact leads to that the evolution of the mean-photon number of the modes
 and the atomic
inversion of the system include information on each other.
In the interaction picture  the unitary evolution operator
of the Hamiltonian (\ref{6}) takes the form
\begin{eqnarray}
\begin{array}{rl}
\hat{U}_{I}(T,0)=\exp(-i\frac{T}{g}\hat{F}_{2})
\\
\\
=\cos (T\hat{D})-i\frac{\sin (T\hat{D})}{g\hat{D}}\hat{F}_{2},\label{8}
\end{array}
\end{eqnarray}
where
\begin{equation}
T=g t,\qquad
\hat{D}^{2}=\hat{a}_{1}^{k_{1}}\hat{a}_{2}^{ k_{2}}\hat{a}_{1}^{\dagger k_{1}}
\hat{a}_{2}^{ \dagger k_{2}}\hat{\sigma}_{+}\hat{\sigma}_{-}+\hat{a}_{1}^{\dagger k_{1}}
\hat{a}_{2}^{ \dagger k_{2}}\hat{a}_{1}^{ k_{1}}\hat{a}_{2}^{ k_{2}}
\hat{\sigma}_{-} \hat{\sigma}_{+}. \label{9}
\end{equation}

For the sake of generalization  we consider
the $j$th mode is initially prepared in the $l$-photon coherent  states
 \cite{jex1,jex2,green} having the form
\begin{equation}
|\psi_{j}(0)\rangle=\sum\limits_{n=0}^{\infty}C^{(j)}_{n}|l_j n\rangle,
\quad
C_n^{(j)}=\exp(-\frac{1}{2}|\alpha_j|^{2})\frac{\alpha_j^n}{\sqrt{n!}}, \label{1}
\end{equation}
where $l_j$ are parameters their values will be specified in the text. Also
throughout  the paper we consider $\alpha_j$ are real.
States (\ref{1}) can be obtained from $l$th harmonic generation using Brandt-Greenberg
operators \cite{green}.
We proceed by considering that  the atom  is initially  in the
 excited state $|+\rangle$.
Therefore, the total initial state of the system is
\begin{equation}
|\Psi (0)\rangle=|\psi_{1} (0)\rangle\bigotimes
|\psi_{2} (0)\rangle\bigotimes
 |+\rangle. \label{3}
\end{equation}
\begin{figure}
 \includegraphics[width=.90\linewidth]{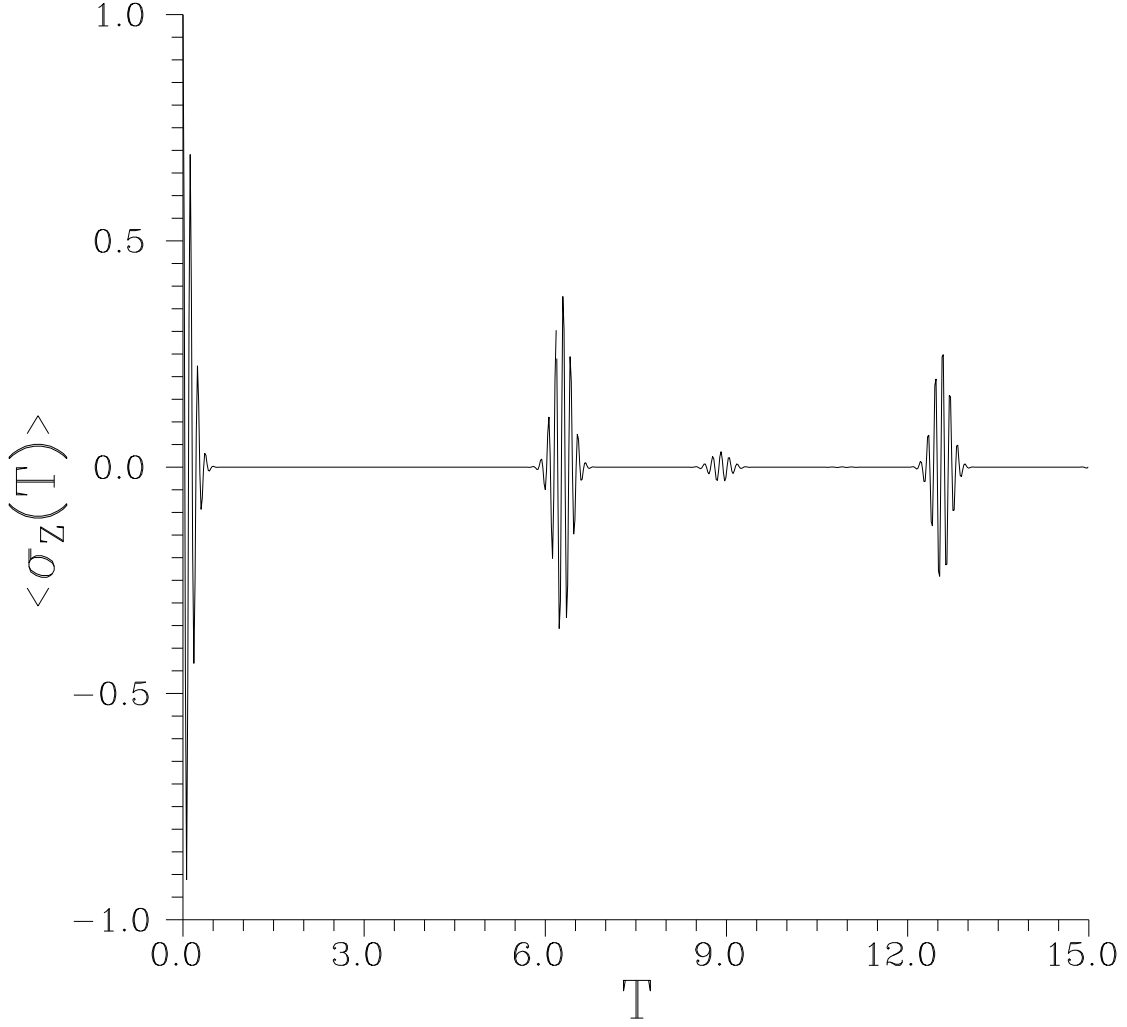}
\caption{ The atomic inversion $\langle\hat{\sigma}_{z}(T)\rangle$
 against the scaled time $T$ when the optical
cavity modes are  initially prepared in the  coherent states
for $(k_{1},k_{2},\alpha_{1},\alpha_{2})=(1,1,5,5)$.
}
\end{figure}

From (\ref{8}) and (\ref{3})
 the dynamical state vector of the system can be evaluated as
\begin{equation}
|\Psi (T)\rangle= \sum\limits_{n,m=0}^{\infty}C_{n,m} \left[
\cos(T\Lambda_{n,m})|+,l_1 n,l_2 m\rangle -i
\sin(T\Lambda_{n,m})|-,l_1 n+k_{1},l_2 m+k_{2}\rangle\right],
 \label{a8a}
 \end{equation}
 where
 $|-\rangle$ denotes ground atomic state,
 $C_{n,m}=C^{(1)}_{n}C^{(2)}_{m}$
and
\begin{equation}
\Lambda_{n,m}=\sqrt{\frac{(l_1 n+k_{1})!(l_2 m+k_{2})!}{(l_1 n)!(l_2 m)!}}.
\label{8b}
\end{equation}

The atomic inversion associated with (\ref{a8a}) is
\begin{equation}
\langle\hat{\sigma}_{z}(T)\rangle=\sum\limits_{n,m}^{\infty}
 C^{2}_{n,m}  \cos(2T\Lambda_{n,m}).
 \label{10a}
\end{equation}

To investigate the evolution of quadrature squeezing we evaluate
the general form for the different moments
of the $\hat{a}_j^{\dagger}$
and $\hat{a}_j$ for (\ref{a8a}) when $l_1=l_2=1$ as
\begin{eqnarray}
\begin{array}{lr} \langle\hat{a}_{1}^{\dagger s'_{1}}(T)
\hat{a}_{1}^{s'_{2}}(T) \hat{a}_{2}^{\dagger s'_{3}}(T)
\hat{a}_{2}^{s'_{4}}(T) \rangle=\sum\limits_{n,m=0}^{\infty}
C_{n+s'_{1},m+s'_{3}}C_{n+s'_{2},m+s'_{4}} \Bigl[
\cos(T\Lambda_{n+s'_{1},m+s'_{3}})\cos(T\Lambda_{n+s'_{2},m+s'_{4}})
 \\
\\
\times\frac{\sqrt{(n+s'_{1})!(n+s'_{2})!(m+s'_{4})!(m+s'_{3})!  }}
{n!m!}\\
 \\
+\sin(T\Lambda_{n+s'_{1},m+s'_{3}})
\sin(T\Lambda_{n+s'_{2},m+s'_{4}})
\frac{\sqrt{(n+k_{1}+s'_{1})!(n+k_{1}+s'_{2})!(m+k_2+s'_{4})!(m+k_2+s'_{3})!  }}
{(n+k_{1})!(m+k_2)!}
\Bigr], \label{12}
\end{array}
\end{eqnarray}
where $s'_{j}, j=1,2,3,4$ are positive integers. Also  we  define
two quadratures $\hat{X}$ and $\hat{Y}$, which denote the real
(electric) and imaginary (magnetic) parts of the radiation field.
Assuming that these quadratures satisfy the following commutation
rule:
\begin{equation}
[\hat{X},\hat{Y}]=\frac{i\hat{d}}{2}, \label{ol1}
\end{equation}
where $\hat{d}$ may be $c$-number or operator. The uncertainty
relation related to the commutation rule (\ref{ol1}) is
\begin{equation}
\langle (\triangle \hat{X})^{2}\rangle \langle (\triangle \hat{Y}
)^{2}\rangle \geq \frac{|\langle \hat{d}\rangle|^{2}}{16},
\label{ol2}
\end{equation}
where  $\langle (\triangle \hat{X})^{2}\rangle=\langle
\hat{X}^{2}\rangle -\langle \hat{X}\rangle ^{2}$ and similar form
can be given for $\langle (\triangle \hat{Y})^{2}\rangle$. The
system is said to be squeezed in the $X$-quadrature if

\begin{equation}
 S(T)=2\langle (\triangle \hat{X}(T))^2\rangle -
\frac{1}{2}|\langle \hat{d}\rangle|\leq 0.
\label{ol3}
\end{equation}
The equality sign in (\ref{ol3}) holds  for minimum-uncertainty
states. Similar definition can be given for the $Y$-quadrature
(defining a $Q$-factor).
As we mentioned in the Introduction we study the evolution of
 four types of quadrature squeezing:
single-mode, two-mode, sum and difference squeezing.
The object of such study is to follow the possible occurrence of the RCP in the
evolution of the squeezing factors  and the conditions required for such
occurrence. Also we try to find which type of squeezing factors can fit
well information on the evolution of the  atomic inversion.
These issues will be discussed  in the following sections.

\section{Single-mode squeezing}
In this section  we study the occurrence of the RCP in the evolution of
the single-mode
squeezing factors for TJCM. The single-mode  squeezing factors
$S_j(t)$ and $Q_j(t)$ for the $j$th-mode when the quadratures $\hat{X}$
and $\hat{Y}$ are defined in the standard form, can be expressed as
\begin{eqnarray}
\begin{array}{lr}
S_{j}(T)=
\langle\hat{a}_{j}^{\dagger}(T)\hat{a}_{j}(T)\rangle+{\rm Re}
\langle\hat{a}_{j}^{2}(T)\rangle-2\Bigl({\rm
Re}\langle\hat{a}_{j}(T)\rangle\Bigr)^{2},
\\
\\
Q_{j}(T)=
\langle\hat{a}_{j}^{\dagger}(T)\hat{a}_{j}(T)\rangle-{\rm Re}
\langle\hat{a}_{j}^{2}(T)\rangle-2\Bigl({\rm
Im}\langle\hat{a}_{j}(T)\rangle\Bigr)^{2}.
\label{s14}
\end{array}
\end{eqnarray}
In the following parts we investigate the natural and numerical
approaches in a greater details.
\subsection{Natural approach}
In this part  and throughout the paper the natural approach is given for
 the standard TJCM. This approach is based on the fact that $\hat{F}_1$ is a constant of
motion and hence the quantities
$\langle\hat{\sigma}_{z}(T)\rangle,
\langle\hat{a}_{1}^{\dagger }(T)\hat{a}_{1}(T)\rangle$ and
$\langle\hat{a}_{2}^{\dagger }(T)\hat{a}_{2}(T)\rangle$
can carry information on each other. Thus the squeezing factors
(\ref{s14}) can give
 information on $\langle\hat{\sigma}_{z}(T)\rangle$ when
\begin{equation}
\langle\hat{a}_{j}(T)\rangle=0,\qquad
\langle\hat{a}_{j}^{2}(T)\rangle=0 \label{s15}
\end{equation}
simultaneously.
This situation can be established when the $j$th  mode
is initially prepared in three-photon states \cite{gener},
 four-photon states \cite{lyn} and so on. Also for particular values
 of the parameter $l$ the $l$-coherent state (\ref{1}) can fulfill  conditions (\ref{s15}).
For such type of initial states
one can  easily prove that
\begin{equation}
\langle\hat{\sigma}_{z}(T)\rangle= 2\langle\hat{a}_{j}^{\dagger
}(0)\hat{a}_{j}(0)\rangle +1 -2S_j(T). \label{ss15}
\end{equation}
Expression (\ref{ss15}) shows that the atomic inversion can be readout
from the quadrature squeezing.
\begin{figure}
  \includegraphics[width=.90\linewidth]{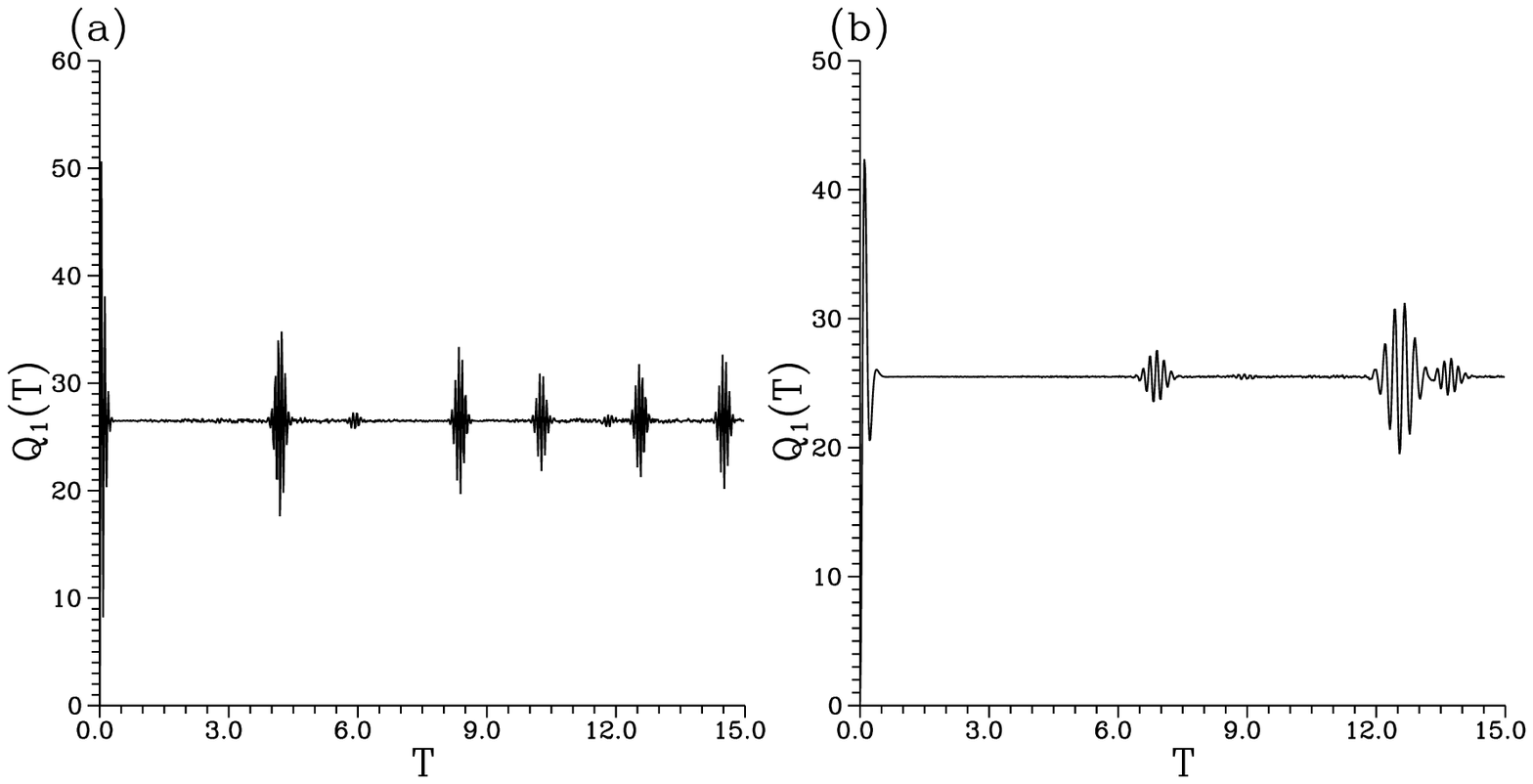}
\caption{ The single-mode squeezing factors against the scaled
time $T$ when the modes are initially prepared in the coherent
light with $(\alpha_{1},\alpha_{2})=(5,5)$ and for
$(k_{1},k_{2})=(3,1)$ (a), (1,3) (b) and (2,2) (c).}
\end{figure}

\subsection{Numerical simulation}
In this part and throughout the paper numerical-simulation approach is
applied to TJCM when $k_1+k_2>2$ and the modes are initially prepared in
coherent light, i.e. $l_1=l_2=1$ in (\ref{1}).
 Now the object here is to discuss the possibility of
obtaining RCP in the evolution of
 the single-mode
squeezing factors  similar to that of the atomic inversion
of the standard TJCM, i.e. $\langle \hat{\sigma}_{z}(T)\rangle_{k_1=k_2=1}$.
The procedures related to this technique are given in \cite{fas1} and we
briefly explain them for the first mode.
From (\ref{s14})   RCP may  occur in $S_{j}(T)$
(or $Q_{j}(T)$)  only when the evolution of the
${\rm Re} \langle \hat{a}_j(T)\rangle$
(or ${\rm Im} \langle \hat{a}_j(T)\rangle$) are close to  zero (, i.e.
steady state) since these quantities are squared. Thus
 when the probability amplitudes
$C_{n,m}$ are real  ${\rm Im} \langle \hat{a}_j(T)\rangle =0$ and
consequently the RCP can likely occur in the evolution of $Q_j(T)$.
Additionally,
when  $k_1+k_2>2$,
$\langle\hat{a}_j^{\dagger}(T)\hat{a}_j(T)\rangle$ exhibits chaotic
behaviour. From numerical data for this case we can
consider $\langle\hat{a}_j^{\dagger}(T)\hat{a}_j(T)\rangle
\simeq\langle\hat{a}_j^{\dagger}(0)\hat{a}_j(0)\rangle$.
This means that the occurrence of the RCP (if it is so) in $Q_j(T)$
is related to the quantity  ${\rm Re} \langle \hat{a}^{2}(T)\rangle$.
Consequently,  we compare the form of
${\rm Re} \langle \hat{a}_j^{2}(T)\rangle$
with that of $\langle \hat{\sigma}_{z}(T)\rangle_{k_1=k_2=1}$.
Now we give a closer look at    $\langle
\hat{a}_1^{2}(T)\rangle$, which from (\ref{12}) has the form:

\begin{eqnarray}
\begin{array}{lr}
\langle \hat{a}_1^{2}(T)\rangle=
\alpha_1^2\sum\limits_{n,m=0}^{\infty}
P(n)P(m)
\Bigl[
\cos(T\Lambda_{n+2,m})\cos(T\Lambda_{n,m})\\
\\
+\sqrt{\frac{(n+k_1+1)(n+k_2+2)}{(n+1)(n+2)}}
\sin(T\Lambda_{n+2,m})\sin(T\Lambda_{n,m})\Bigr],
\label{s14s}
\end{array}
\end{eqnarray}
where $P(n)=(C^{(1)}_n)^2$ is the photon-number distribution for the coherent light.
In the  strong-intensity regime, i.e. $\bar{n}_j=\langle
\hat{a}_j^{\dagger}(0)\hat{a}_j(0)\rangle =|\alpha_j|^2 >>1$, and
 finite values of
the transition parameters $k_j$
 we can apply the harmonic approximation technique
\cite{eber,chr}. This technique is based on the fact that the
photon-number distribution  of the coherent light is Poissonian
with a sharp peak at $n=\bar{n}$ and hence the terms which
contribute
 effectively to the summation in (\ref{s14s}) are those for which
 $ n\simeq\bar{n}$.
As a result of this fact
 the  square root in the second line of (\ref{s14s})
 tends to unity and (\ref{s14s}) reduces to

\begin{equation}
\langle \hat{a}_1^{2}(T)\rangle\simeq
\bar{n}_1\sum\limits_{n,m=0}^{\infty}
P(n)P(m)
\cos[T(\Lambda_{n+2,m}-\Lambda_{n,m})].
\label{1s4s}
\end{equation}
Regardless of the prefactor $\bar{n}_1$
 in (\ref{1s4s}), the comparison between (\ref{10a}) of the case
 $(k_j,l_j)=(1,1)$ and (\ref{1s4s}) leads to that both
expressions exhibit quite similar dynamical behaviour only
when the arguments
of cosines in the two expressions are comparable. Therefore, we seek
 the proportionality factor $\mu_1$, say, which can be evaluated from the
following expression
\begin{equation}
\mu_1=\frac{
\Lambda_{n+2,m}-\Lambda_{n,m}}{2\sqrt{(n+1)(m+1)}}.
\label{pro1}
\end{equation}
Expression (\ref{pro1}) can be re-expressed  as
\begin{eqnarray}
\begin{array}{lr}
\mu_1=\sqrt{\frac{(m+k_2)!(n+k_1)!}{(n+1)!(m+1)!}}
\Bigl\{\frac{(2n+3)k_1+k_1^2}{2\sqrt{(n+2)(n+1)}
[\sqrt{(n+k_1+2)(n+k_1+1)}+
\sqrt{(n+2)(n+1)}]}\Bigr\}\\
\\
=\frac{n^{\frac{k_1-3}{2}}m^{\frac{k_2-1}{2}}}{2}
\left[\prod\limits_{j=0}^{k_1}(1+\frac{k_1-j}{n})\right]^{\frac{1}{2}}
\left[\prod\limits_{j'=0}^{k_2}(1+\frac{k_2-j'}{m})\right]^{\frac{1}{2}}\\
\\
\times\frac{(2+\frac{3}{n})k_1+\frac{k_1^2}{n}}{(1+\frac{1}{n})
\sqrt{(1+\frac{1}{n})
(1+\frac{1}{m})}[
\sqrt{(1+\frac{k_1+2}{n})(1+\frac{k_1+1}{n})}+
\sqrt{(1+\frac{2}{n})(1+\frac{1}{n})}]}.
\label{pro2}
\end{array}
\end{eqnarray}
\begin{figure}
  \includegraphics[width=.90\linewidth]{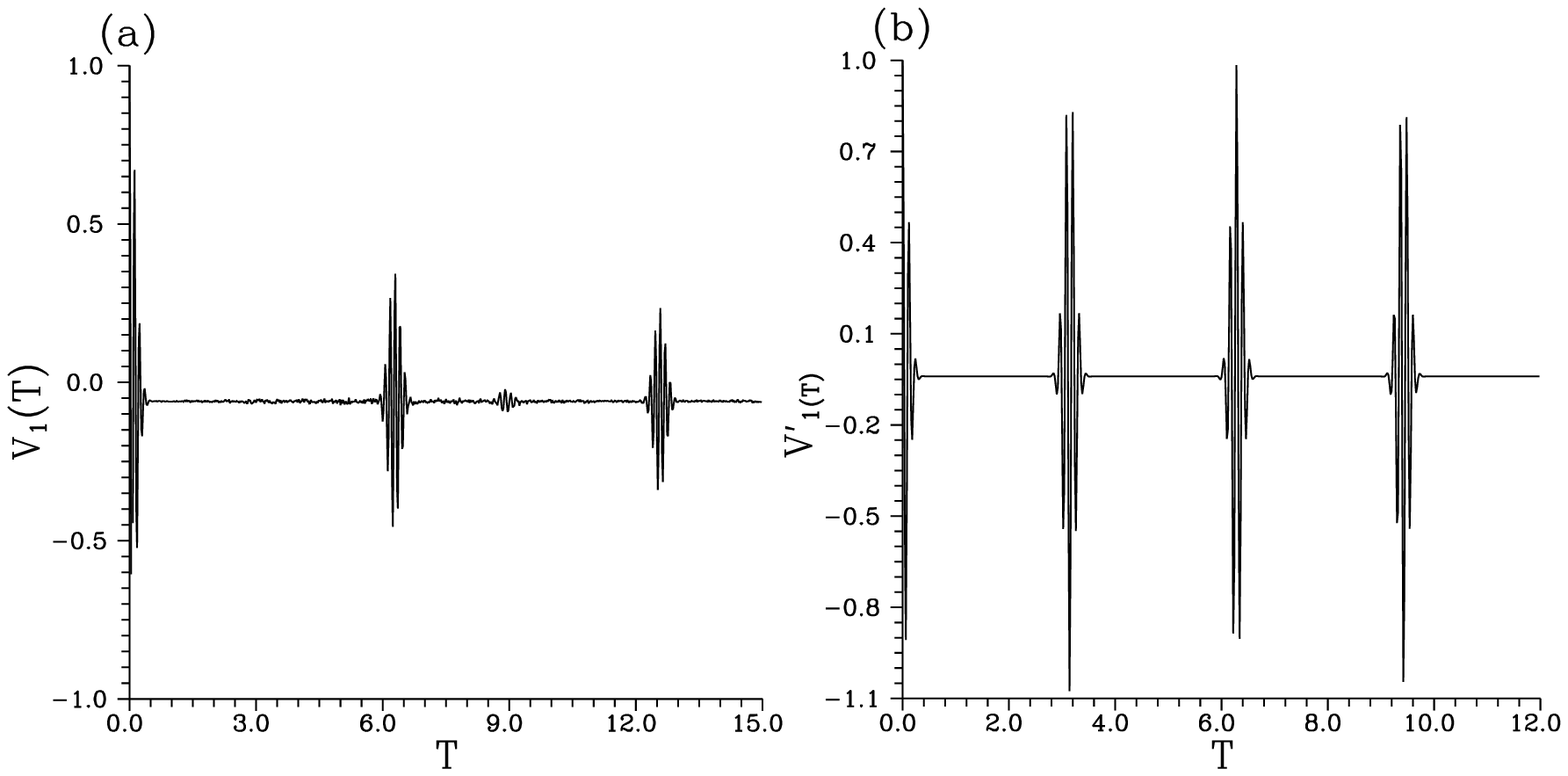}
\caption{ The modified squeezing factor for the cases
$(k_1,k_2)=(3,1)$ (a) and $(k_1,k_2)=(2,2)$ (b) when the modes are
initially prepared in
 the coherent light with $\alpha_1=\alpha_2=5$.}
\end{figure}
In the framework of  the harmonic approximation (i.e.
$\epsilon/\bar{n}\rightarrow 0$ where $\epsilon$ is an arbitrary finite
number)  the second
part of (\ref{pro2})  reduces to
\begin{equation}
\mu_1\simeq \frac{k_1}{2}\bar{n}^{\frac{k_1-3}{2}}
\bar{m}^{\frac{k_2-1}{2}}. \label{pro3}
\end{equation}
Expression (\ref{pro3}) shows that there are three cases can provide RCP
in the evolution of the $Q_1(T)$, which are: $(k_1,k_2,\mu_1)=(3,1,3/2)$
and $(k_1,k_2,\mu_1)=(1,3,1/2), (2,2,1)$. For the latter cases
 the values of $\bar{n}$ and $\bar{m}$ have to be comparable.
Information about these cases has been shown in Figs. 2(a)--(c) for given
values of the parameters. It is obvious that we have three different shapes
of the RCP.
Form Fig. 2(a) revivals and secondary revivals are remarkable having
shapes similar to those of the
$\langle \hat{\sigma}_{z}(T)\rangle_{k_1=k_2=1}$ (compare to Fig. 1).
 Fig. 2(b) includes  revivals and collapses
but their forms are different  form those
of the $\langle \hat{\sigma}_{z}(T)\rangle_{k_1=k_2=1}$.
Nevertheless, in Fig. 2(c) the RCP is systematic, i.e. revivals are compact and
occur periodically with period $\pi$. Actually, this behaviour is quite
typical with the evolution of the
atomic inversion  of the single-mode two-photon JCM, i.e.
$\langle \hat{\sigma}_{z}(T)\rangle_{k_1=2,k_2=0}$.
This can be analytically realized by applying the harmonic approximation
to the argument of the cosine in (\ref{1s4s}) for the case
$(k_1,k_2)=(2,2)$ as
\begin{eqnarray}
\begin{array}{lr}
\Lambda_{n+2,m}-\Lambda_{n,m}=\sqrt{(m+1)(m+2)}
\left[\sqrt{(n+3)(n+4)}-\sqrt{(n+1)(n+2)}\right]\\
\\
\simeq\sqrt{(m+1)(m+2)}[\bar{n}+ \frac{7}{2}-\bar{n}-\frac{3}{2}]\\
\\
=2\sqrt{(m+1)(m+2)}.
\label{proo3}
\end{array}
\end{eqnarray}
The last line in  (\ref{proo3}) is typical the argument of
the cosine of the
$\langle \hat{\sigma}_{z}(T)\rangle_{k_1=2, k_2=0}$, e.g. see equation
(12) in \cite{amita} for $\chi=0$.

Now we deduce the rescaled squeezing factors for the cases $(k_1,k_2)=(3,1)$
and $(2,2)$, which can give
$\langle \hat{\sigma}_{z}(T)\rangle_{k_1=k_2=1}$
and $\langle \hat{\sigma}_{z}(T)\rangle_{k_1=2,k_2=0}$, respectively.
 From Fig. 2(a) and the discussion given above we can write the rescaled
squeezing factor for $(k_1,k_2)=(3,1)$ as
\begin{equation}
V_{1}(T)=\frac{\langle\hat{n}_1(0)\rangle
-Q_{1}(\frac{2}{3}T)}
{\langle\hat{n}_1(0)\rangle}. \label{29}
\end{equation}

Similarly the rescaled squeezing factor for the case $(k_1,k_2)=(2,2)$
is
\begin{equation}
V'_1(T)=
\frac{\langle\hat{n}_1(0)\rangle
-Q_{1}(T)}
{\langle\hat{n}_1(0)\rangle}.
\label{pro4}
\end{equation}
Expression (\ref{29}) and (\ref{pro4})  have  been depicted in Figs. 3(a)
and (b) for the  values
of the interaction parameters as those given for Figs. 2(a) and (b), respectively.
 Comparison between
Fig. 1 and Fig. 3(a) demonstrates our conclusion.
Also Fig. 3(b) provides completely typical shape as that of
$\langle \hat{\sigma}_{z}(T)\rangle_{k_1=2, k_2=0}$ (see Fig. 2 in
\cite{jos}).

\section{Two-mode  squeezing}
In this section, we use
procedures similar to those given in section 3 to investigate the RCP in
the evolution of the two-mode squeezing for TJCM.
Starting with the two-mode squeezing factors, which can be expressed as
\begin{eqnarray}
\begin{array}{lr}
S_{2}(T)=
{\rm Re}\langle
\hat{a}_{2}^{\dagger}(T)\hat{a}_{1}(T)
+\hat{a}_{1}(T)\hat{a}_{2}(T)\rangle-
2{\rm Re}\langle\hat{a}_{1}(T)\rangle
{\rm Re}\langle\hat{a}_{2}(T)\rangle
 \\
 \\
 +\sum\limits_{j=1}^{2}
\Bigl[{\rm Re}\langle\hat{a}_{j}^{\dagger}(T)\hat{a}_{j}(T)\rangle+{\rm Re}
\langle\hat{a}_{j}^{2}(T)\rangle-2\Bigl({\rm
Re}\langle\hat{a}_{j}(T)\rangle\Bigr)^{2}\Bigr],
\\
\\
Q_{2}(T)
=
{\rm Re}\langle
\hat{a}_{2}^{\dagger}(T)\hat{a}_{1}(T)
-\hat{a}_{1}(T)\hat{a}_{2}(T)\rangle-
2{\rm Im}\langle\hat{a}_{1}(T)\rangle
{\rm Im}\langle\hat{a}_{2}(T)\rangle
 \\
 \\
 +\sum\limits_{j=1}^{2}
\Bigl[\langle\hat{a}_{j}^{\dagger}(T)\hat{a}_{j}(T)\rangle-{\rm Re}
\langle\hat{a}_{j}^{2}(T)\rangle-2\Bigl({\rm
Im}\langle\hat{a}_{j}(T)\rangle\Bigr)^{2}\Bigr].
\label{14}
\end{array}
\end{eqnarray}
We discuss the natural and numerical approaches for two-mode squeezing
in the following parts.
\subsection{Natural phenomenon}
Explanations similar to those given in subsection 3.1
the two-mode squeezing factors can give direct information on the corresponding
$\langle\hat{\sigma}_{z}(T)\rangle$
for  initial states, which satisfy simultaneously
the following conditions:
\begin{equation}
\langle\hat{a}_{j}(T)\rangle=0,\qquad
\langle\hat{a}_{j}^{2}(T)\rangle=0,\qquad  j=1,2 \label{15}
\end{equation}
We should stress that (\ref{15}) requires the two modes to be initially
prepared in such type of states. This is different from (\ref{s15})
of the single-mode
squeezing, which requires the mode under consideration only to be initially
in such states.
In this case the two squeezing factors (\ref{14})
are typical and can give $\langle \hat{\sigma}_{z}(T)\rangle$.
 Similar to the single-mode squeezing case this can be verified
 when the two modes are initially prepared in
the $l$-photon coherent states with $l=3,4,..$, etc.
For such states  one can easily prove that
\begin{equation}
\langle\hat{\sigma}_{z}(T)\rangle= \langle\hat{a}_{1}^{\dagger
}(0)\hat{a}_{1}(0)\rangle + \langle\hat{a}_{2}^{\dagger
}(0)\hat{a}_{2}(0)\rangle+1-S_2(T). \label{ss15a}
\end{equation}

\subsection{Numerical simulation}
Discussion similar to that  given in subsection 3.2
one can easily prove that RCP can  occur in $Q_{2}(T)$ when
 $(k_1,k_2)=(3,1), (1,3)$ and $(2,2)$.
Also  one can easily realize
for the case $(k_1,k_2)=(1,3)$ that $Q_2(T)$ exhibits
 RCP, which is the combination from those shown in Figs. 2(a) and (b), i.e.
RCP is different from that  of the
$\langle \hat{\sigma}_{z}(T)\rangle_{k_1=1, k_2=1}$. Nevertheless, for the
case $(k_1,k_2)=(2,2)$, $Q_2(T)$ can give information on the
$\langle \hat{\sigma}_{z}(T)\rangle_{k_1=2, k_2=0}$.
The rescaled squeezing factor, which is typical
      $\langle \hat{\sigma}_{z}(T)\rangle_{k_1=2, k_2=0}$,
      can be evaluated as
\begin{equation}
V'_2(T)=\frac{\langle\hat{n}_1(0)\rangle +\langle\hat{n}_2(0)\rangle
-Q_{2}(T)}
{\langle\hat{n}_1(0)\rangle +\langle\hat{n}_2(0)\rangle  }.
\label{pro4o}
\end{equation}

\section{Sum and difference squeezing}
In this section we investigate the occurrence of the RCP in the evolution
of the sum and difference squeezing factors \cite{hil}. In these factors  the
intermode correlation is involved in the quadrature squeezing.
We have noted for numerical-simulation approach that the technique
given in section 3 is
partially working for sum and difference squeezing. More illustratively, it can give the exact values for
the transition parameters $k_j$ whose squeezing factors exhibit RCP
but it fails to provide the correct rescaled  squeezing factor.
 Nevertheless, this difficulty can be numerically treated.
We have noted that
the sum and difference squeezing give
 only information on the occurrence of the revivals (not secondary
revivals) in $\langle \hat{\sigma}_{z}(T)\rangle_{k_1=1, k_2=1}$.
Moreover, sum squeezing can provide
$\langle \hat{\sigma}_{z}(T)\rangle_{k_1=2, k_2=0}$, however,
difference squeezing fails.
We discuss all these results in the following.
For the sum squeezing we have \cite{hil}
\begin{equation}
\hat{X}=\frac{1}{2}[\hat{a}_{1}\hat{a}_{2}+\hat{a}^{\dagger}_{1}\hat{a}^{\dagger}_{2}],
\quad
\hat{Y}=\frac{i}{2}[\hat{a}^{\dagger}_{1}\hat{a}^{\dagger}_{2}-
\hat{a}_{1}\hat{a}_{2}],\quad \hat{C}=\hat{a}_{1}^{\dagger }
\hat{a}_{1}+\hat{a}_{2}^{\dagger } \hat{a}_{2} +1.
 \label{30}
\end{equation}
Therefore, sum-squeezing factors can be expressed as
\begin{eqnarray}
\begin{array}{lr} S_{3}(T)={\rm
Re}\langle\hat{a}_{1}^{2}(T)\hat{a}_{2}^{2}(T)\rangle + \langle
\hat{a}_{1}^{\dagger }(T) \hat{a}_{1}(T) \hat{a}_{2}^{\dagger }(T)
\hat{a}_{2}(T)\rangle -2\Bigl({\rm Re}\langle
\hat{a}_{1}(T)\hat{a}_{2}(T) \rangle \Bigr)^{2},
\\
\\
Q_{3}(T)=
 \langle \hat{a}_{1}^{\dagger }(T) \hat{a}_{1}(T)
\hat{a}_{2}^{\dagger }(T) \hat{a}_{2}(T)\rangle
-{\rm Re} \langle\hat{a}_{1}^{2}(T)\hat{a}_{2}^{2}(T)\rangle
-2\Bigl({\rm Im}\langle \hat{a}_{1}(T)\hat{a}_{2}(T) \rangle\Bigr)^{2}.
\label{31}
\end{array}
\end{eqnarray}

For difference squeezing the quadratures $\hat{X}$ and $\hat{Y}$
can be obtain from  those in
(\ref{30}) by using the transformation $\hat{a}_1\leftrightarrow
\hat{a}_1^{\dagger}$ and consequently $\hat{C}=
\hat{a}_2^{\dagger}\hat{a}_2-\hat{a}_1^{\dagger}\hat{a}_1$. For the sake
of simplicity we assume that the two modes are initially prepared in states
having the
same photon-number distribution with $\alpha_1=\alpha_2$. This leads to that
$\langle\hat{a}_2^{\dagger}(T)\hat{a}_2(T)\rangle =
\langle\hat{a}_1^{\dagger}(T)\hat{a}_1(T)\rangle$, i.e. $|
\langle \hat{C}(T)\rangle|=0$.
Under these conditions the difference squeezing factors take the forms
\begin{eqnarray}
\begin{array}{lr} S_{4}(T)= {\rm Re}\langle\hat{a}_{1}^{\dagger
2}(T)\hat{a}_{2}^{2}(T)\rangle + \langle \hat{a}_{1}^{\dagger }(T)
\hat{a}_{1}(T) \hat{a}_{2}^{\dagger }(T) \hat{a}_{2}(T)\rangle +
\langle \hat{a}_{1}^{\dagger }(T) \hat{a}_{1}(T)\rangle
-2\Bigl({\rm Re}\langle \hat{a}_{1}(T)\hat{a}_{2}^{\dagger }(T)
\rangle \Bigr)^{2},
\\
\\
Q_{4}(T)=
 \langle \hat{a}_{1}^{\dagger }(T) \hat{a}_{1}(T)
\hat{a}_{2}^{\dagger }(T) \hat{a}_{2}(T)\rangle
+\langle \hat{a}_{1}^{\dagger }(T) \hat{a}_{1}(T)\rangle
-{\rm Re} \langle\hat{a}_{1}^{\dagger 2}(T)\hat{a}_{2}^{2}(T)\rangle
-2\Bigl({\rm Im}\langle \hat{a}_{1}(T)\hat{a}_{2}^{\dagger }(T) \rangle
\Bigr)^{2}.
\label{31d}
\end{array}
\end{eqnarray}
\begin{figure}
 \includegraphics[width=.90\linewidth]{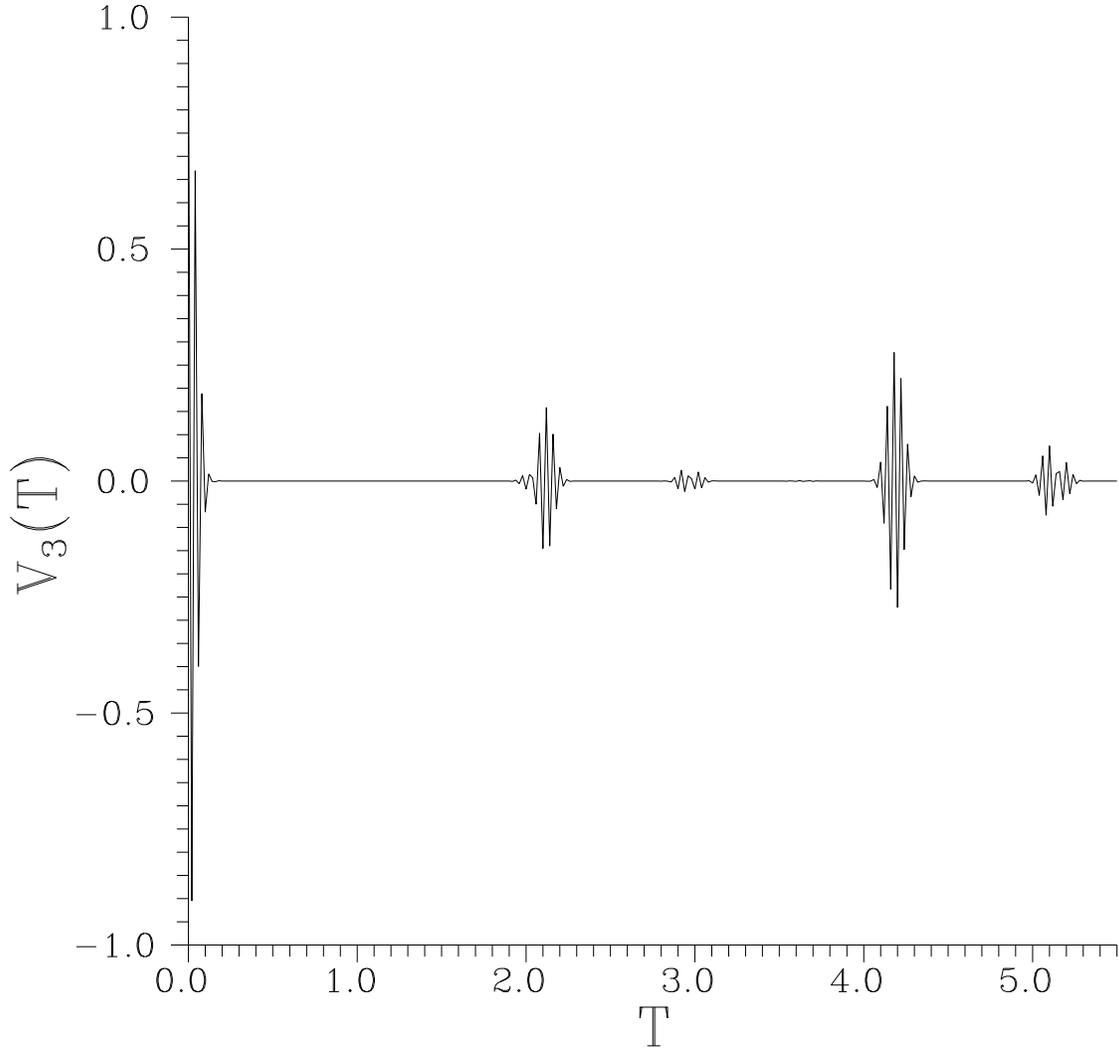}
\caption{ The rescaled squeezing factor $V_3(T)$
 against the scaled time $T$ when the optical
cavity modes are  initially prepared in the  three-photon coherent states
and $(k_{1},k_{2},\alpha_{1},\alpha_{2})=(1,1,5,5)$.}
\end{figure}

\subsection{Natural phenomenon}
Now  we  seek states that evolve with the TJCM causing the evolution  of the
$\langle\hat{a}_{1}^{ 2}(T)\hat{a}_{2}^{ 2}(T)\rangle$  and
$\langle \hat{a}_{1}(T)\hat{a}_{2}(T) \rangle$ close to zero.
 This can occur if one of the two modes at least is initially in, e.g.,
the three-photon or four-photon states, (cf. (\ref{1})).
 For such  states expressions (\ref{31}) of sum squeezing reduce to
\begin{equation}
Q_{3}(T)=S_{3}(T)=
 \langle \hat{a}_{1}^{\dagger }(T) \hat{a}_{1}(T)
\hat{a}_{2}^{\dagger }(T) \hat{a}_{2}(T)\rangle.
\label{31a}
\end{equation}
In the framework of harmonic approximation the rescaled squeezing factor
associated with (\ref{31a}), which  can provide the corresponding atomic inversion, is
\begin{equation}
V_3(T)\simeq
\frac{2\langle\hat{n}_{1}(0)\rangle\langle\hat{n}_{2}(0)\rangle+
\langle\hat{n}_{1}(0)\rangle+\langle\hat{n}_{2}(0)\rangle+1-2S_3(T)}
{\langle\hat{n}_{1}(0)\rangle+\langle\hat{n}_{1}(0)\rangle +1}.
\label{35}
\end{equation}
In Fig. 4 we have plotted (\ref{35})
when the modes are initially
prepared in the three-photon coherent states. Actually, we have found
that $V_3(T)=\langle \hat{\sigma}_{z}(T)\rangle$.
 From Fig. 4 the revivals and secondary revivals are remarkable.
 Additionally, the revival times of this case
are three times smaller than those of the initial coherent light
since we are dealing with three-photon  states (compare Fig. 1 and Fig. 4).

Similarly for the difference squeezing (\ref{31d}) we can obtain
\begin{equation}
\langle\hat{\sigma}_z(T)\rangle\simeq
\frac{
\left(\langle\hat{n}_{1}(0)\rangle+1\right)^2
-Q_4(T)}
{\langle\hat{n}_{1}(0)\rangle+1}.
\label{35d}
\end{equation}
\begin{figure}
 \includegraphics[width=.90\linewidth]{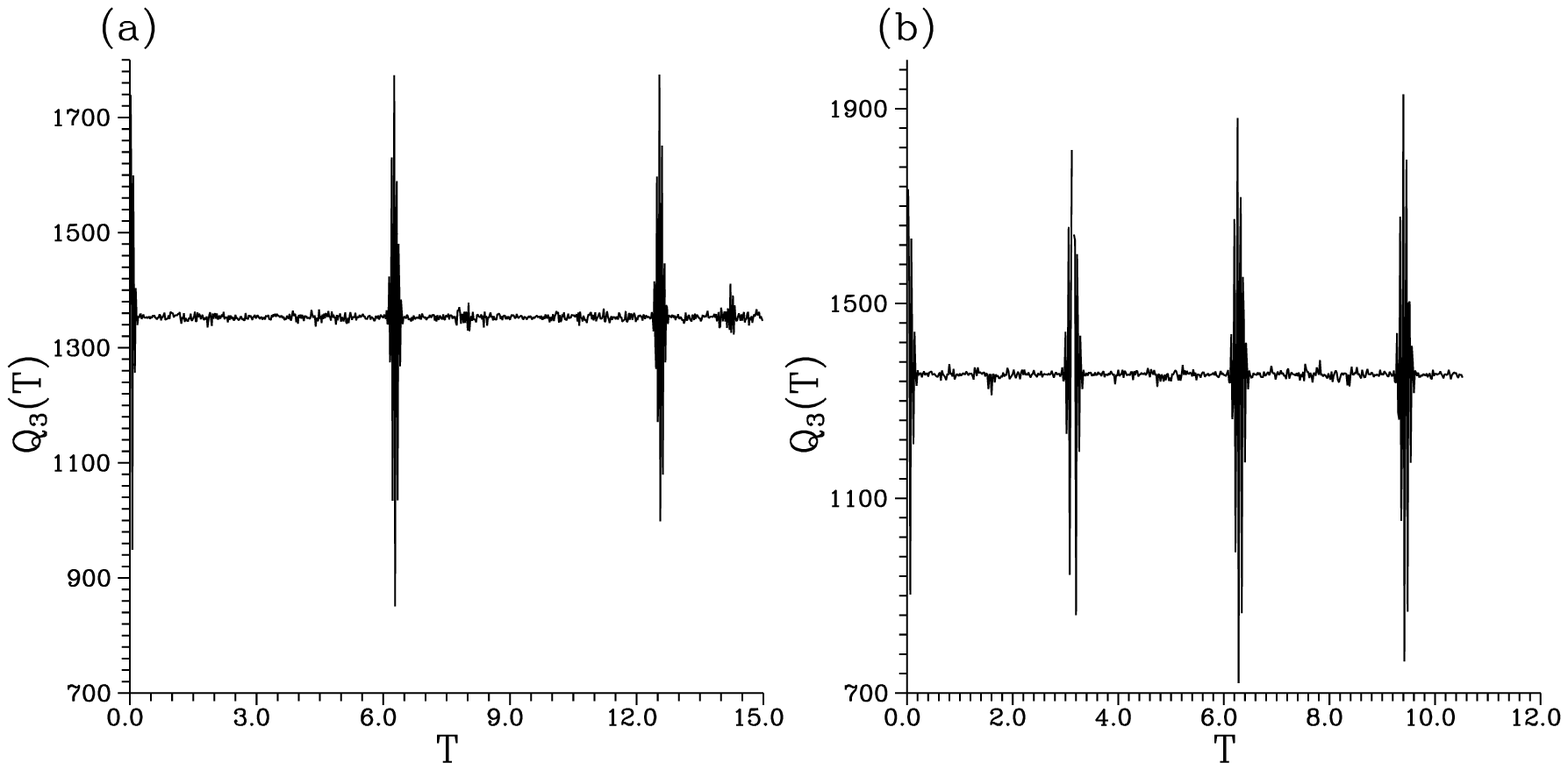}
\caption{The sum-squeezing factor
 against the scaled time $T$ when the optical
cavity modes are  initially prepared in the  coherent states with $\alpha_1=
\alpha_2=5$ for $(k_{1},k_{2})=(3,1)$ (a) and $(2,2)$ (b).}
\end{figure}

\subsection{Numerical simulation}
 Similar arguments
as those given in subsection 3.2 lead to that
if sum-squeezing factor exhibits RCP the quantity
${\rm Re}\langle\hat{a}_1^{2}(T)\hat{a}_2^{2}(T)\rangle$
is responsible for this. From (\ref{12}) and the harmonic approximation
technique we arrive at
\begin{equation}
\langle\hat{a}_1^{2}(T)\hat{a}_2^{2}(T)\rangle\simeq
\alpha_1^2\alpha_2^2\sum\limits_{n,m=0}^{\infty}
C_{n,m}^{2}\cos[T(\Lambda_{n+2,m+2}-\Lambda_{n,m})]. \label{36}
 \end{equation}
Now we seek the proportionality factor, which can be obtained from the
following

\begin{equation}
\mu_2=\frac{
\Lambda_{n+2,m+2}-\Lambda_{n,m}}{2\sqrt{(n+1)(m+1)}}.
\label{pro5}
\end{equation}
After lengthy calculation but straightforward
(\ref{pro5}) reduces to
\begin{eqnarray}
\begin{array}{lr}
\mu_2\simeq \frac{1}{4}\Bigl(2
k_2\bar{n}_1^{\frac{k_1-1}{2}}\bar{n}_2^{\frac{k_2-3}{2}}
+k_2^2\bar{n}_1^{\frac{k_1-1}{2}}\bar{n}_2^{\frac{k_2-5}{2}}
+2k_1\bar{n}_1^{\frac{k_1-3}{2}}\bar{n}_2^{\frac{k_2-1}{2}}
+4k_1k_2\bar{n}_1^{\frac{k_1-3}{2}}\bar{n}_2^{\frac{k_2-3}{2}}\\
\\
+2k_1k_2^2\bar{n}_1^{\frac{k_1-3}{2}}\bar{n}_2^{\frac{k_2-5}{2}}
+k_1^2\bar{n}_1^{\frac{k_1-5}{2}}\bar{n}_2^{\frac{k_2-1}{2}}
+2k_1^2k_2\bar{n}_1^{\frac{k_1-5}{2}}\bar{n}_2^{\frac{k_2-3}{2}}
+k_1^2k_2^2\bar{n}_1^{\frac{k_1-5}{2}}\bar{n}_2^{\frac{k_2-5}{2}}\Bigr).
\label{pro6}
\end{array}
\end{eqnarray}
Similar to the single-mode squeezing case  when $\bar{n}_1\simeq\bar{n}_2$ there are three
cases,
which  can provide RCP in the evolution of the $Q_3(T)$, namely,
$(k_1,k_2)=(3,1), (1,3)$ and $(2,2)$. For these cases the proportionality factor
is $\mu_2=2$. Actually, this factor cannot give the correct
 rescaled squeezing factor.
This fact can be realized from Figs. 5(a) and (b), in which
 we have plotted the sum squeezing factors for
 $(k_1,k_2)=(3,1)$ and $(k_1,k_2)=(2,2)$, respectively.
  From these figures one can see that for the case
 $(k_1,k_2)=(3,1)$ sum squeezing can give in principle information
 on $\langle \hat{\sigma}_{z}(T)\rangle_{k_1=k_2=1}$, however, for
 $(k_1,k_2)=(2,2)$ it can provide the evolution of the
 $\langle \hat{\sigma}_{z}(T)\rangle_{k_1=2,k_2=0}$.
 From Figs. 5 and expression (\ref{31}) the rescaled
squeezing factor is
\begin{equation}
V_4(T)=
\frac{
\langle\hat{n}_{1}(0)\rangle\langle\hat{n}_{2}(0)\rangle
-Q_3(T)}
{\langle\hat{n}_{1}(0)\rangle\langle\hat{n}_{2}(0)\rangle}.
\label{pr35}
\end{equation}
Expression (\ref{pr35})
 gives $\langle \hat{\sigma}_{z}(T)\rangle_{k_1=k_2=1}$
 and $\langle \hat{\sigma}_{z}(T)\rangle_{k_1=2,k_2=0}$
for $(k_1,k_2)=(3,1)$ and $(2,2)$, respectively.

\begin{figure}
  \includegraphics[width=.90\linewidth]{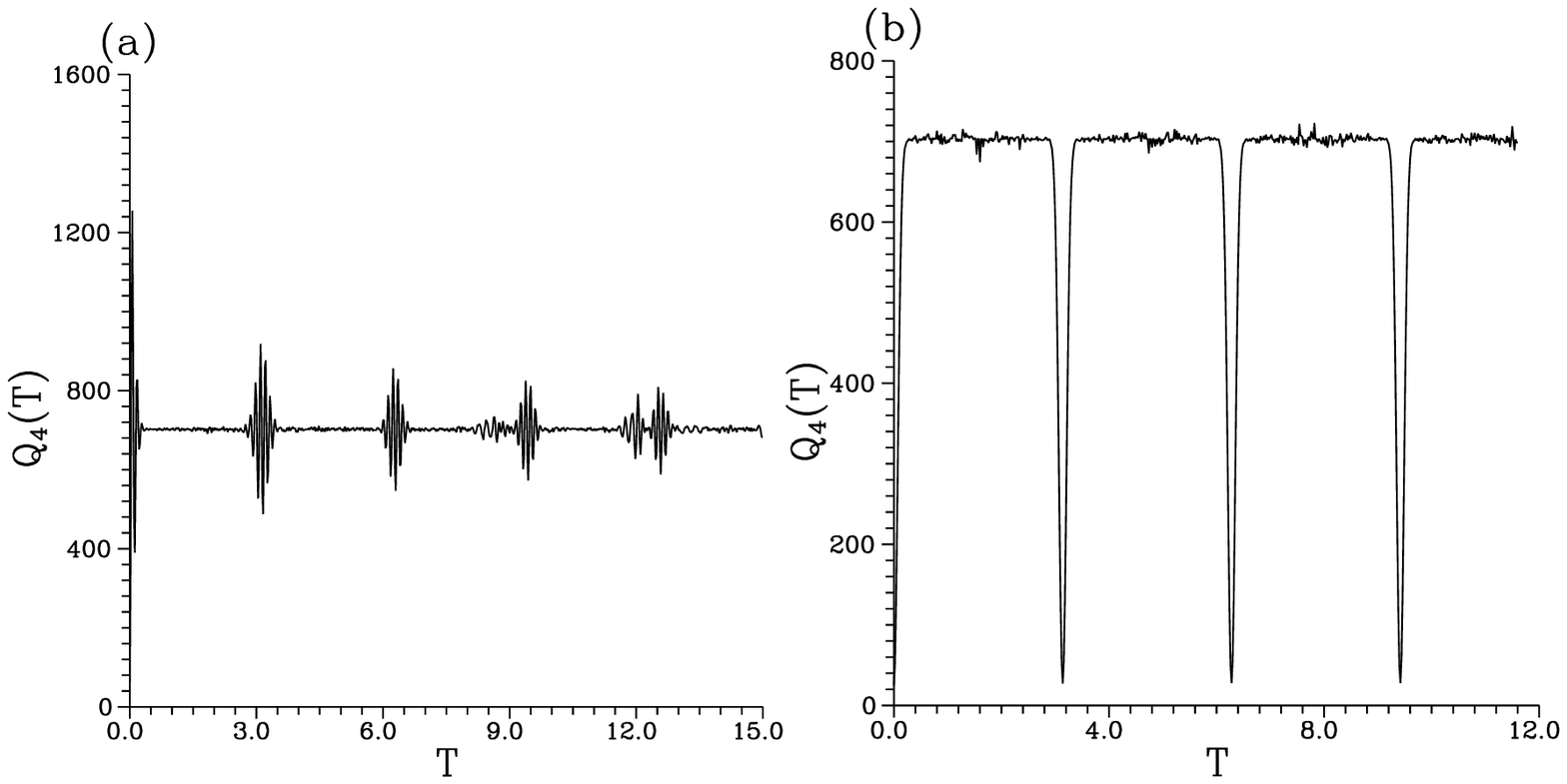}
\caption{ The difference-squeezing factor
 against the scaled time $T$ when the optical
cavity modes are  initially prepared in the  coherent states with $\alpha_1=
\alpha_2=5$
for $(k_{1},k_{2})=(3,1)$ (a) and $(2,2)$ (b).
}
\end{figure}

On the other hand, for the difference squeezing we found that
 the RCP can be remarked in
the evolution of the $Q_4(T)$ when $(k_1,k_2)=(3,1), (1,3)$,
however, for
$(k_1,k_2)=(2,2)$ the technique fails.
To demonstrate these cases we have  plotted
Figs. 6(a) and (b) for $Q_4(T)$ when $(k_1,k_2)=(3,1)$ and $(2,2)$,
respectively. From Fig. 6(a) RCP is established but it is completely different
from that of the
$\langle \hat{\sigma}_{z}(T)\rangle_{k_1=k_2=1}$,
however, Fig. 6(b) exhibits periodically inverted peaks, i.e.
it does not exhibit RCP.
Now from Fig. 6(a) and expression (\ref{31d}), the rescaled
squeezing factor is
\begin{equation}
V_5(T)=
\frac{
\langle\hat{n}_{1}(0)\rangle(\langle\hat{n}_{1}(0)\rangle+1)
-Q_4(\frac{T}{2})}{\langle\hat{n}_{1}(0)\rangle^{2}}.
\label{prd35}
\end{equation}
We have numerically checked (\ref{pr35}) and
(\ref{prd35}) for the case $(k_1,k_2)=(3,1)$
and found that they give typical
behaviour as that of
the $\langle \hat{\sigma}_{z}(T)\rangle_{k_1=k_2=1}$
 except that the secondary revivals are absent.
 Additionally, the widths of the revival patterns of (\ref{prd35})
  are a little bit greater than those
of the $\langle \hat{\sigma}_{z}(T)\rangle_{k_1=k_2=1}$.

\section{Conclusions}
In this paper we have discussed the possibility of including
the squeezing factors of the two-mode multiphoton
 JCM information on the atomic inversion of the standard TJCM.
In contrast to  the single-mode JCM \cite{faisal2} we have various types of
quadrature squeezing, namely, single-mode, two-mode, sum and
difference squeezing.
Two approaches have been applied for
 all these types, which are natural phenomenon and numerical simulation.
Natural approach has been devoted to the standard TJCM and found that there
is a class of states their squeezing factors provide
 the corresponding atomic inversion.
For the numerical-simulation approach
 we have shown that for specific value of the transition parameters,
in particular, $k_1+k_2=4$ the $Y$-quadrature squeezing factor
 can provide
 RCP similar to that associated with  the
$\langle \hat{\sigma}_{z}(T)\rangle_{k_1=k_2=1}$ or
$\langle \hat{\sigma}_{z}(T)\rangle_{k_1=2, k_2=0}$ based on the
values of $k_j$.
Specifically, for $(k_1,k_2)=(3,1)$ single-mode
squeezing factor can give information on
$\langle \hat{\sigma}_{z}(T)\rangle_{k_1=k_2=1}$, however,
 sum and difference squeezing factors   give partial information.
On the other hand, for $(k_1,k_2)=(2,2)$ single-mode, two-mode
 and sum squeezing factors give information on the
 $\langle \hat{\sigma}_{z}(T)\rangle_{k_1=2, k_2=0}$.
Also we have deduced the rescaled-squeezing factors for all these types
 giving  information on the atomic inversion.

We conclude this paper by mentioning that
the influence of the values of the atomic phases on the phenomenon under
consideration is the same as that for the single-mode JCM \cite{faisal2}.
In other words, for natural (numerical)
approach the squeezing factor is sensitive (insensitive)
to the initial atomic state, i.e. natural approach  can provide
"coherent trapping"  \cite{zaheer}. Also for numerical-simulation
approach we have numerically checked
that the RCP occurs in the quadrature squeezing only for $k_1+k_2=4$.

\section*{References}

\end{document}